\documentclass[aip,apl,numerical,reprint,floatfix]{revtex4-1}

\usepackage{amsmath}
\usepackage{units}
\usepackage{graphicx}
\begin{document}


\title{Exchange magnon induced resistance asymmetry in permalloy spin-Hall oscillators} 



\author{S. Langenfeld}
\altaffiliation{current address: Max-Planck-Institut f\"ur Quantenoptik, 85748 Garching, Germany}
\affiliation{Microelectronics Group, Cavendish Laboratory, University of Cambridge, Cambridge, CB3 0HE, UK}
\affiliation{Walter Schottky Institut and Physik-Department, Technische Universit\"at M\"unchen, 85748 Garching, Germany}

\author{V. Tshitoyan}
\affiliation{Microelectronics Group, Cavendish Laboratory, University of Cambridge, Cambridge, CB3 0HE, UK}

\author{Z. Fang}
\affiliation{Microelectronics Group, Cavendish Laboratory, University of Cambridge, Cambridge, CB3 0HE, UK}

\author{A. Wells}
\affiliation{School of Physics and Astronomy, University of Leeds, Leeds, LS2 9JT, UK}

\author{T. A. Moore}
\affiliation{School of Physics and Astronomy, University of Leeds, Leeds, LS2 9JT, UK}

\author{A. J. Ferguson}
\email[]{ajf1006@cam.ac.uk}
\affiliation{Microelectronics Group, Cavendish Laboratory, University of Cambridge, Cambridge, CB3 0HE, UK}


\date{\today}

\begin{abstract}
We investigate magnetization dynamics in a spin-Hall oscillator using a direct current measurement as well as conventional microwave spectrum analysis. When the current applies an anti-damping spin-transfer torque, we observe a change in resistance which we ascribe to the excitation of incoherent exchange magnons. 
A simple model is developed based on the reduction of the effective saturation magnetization, quantitatively explaining the data. 
The observed phenomena highlight the importance of exchange magnons on the operation of spin-Hall oscillators.
\end{abstract}


\maketitle 

The combination of the spin-Hall effect and the spin-transfer torque may be used to compensate the magnetic damping of a ferromagnet, facilitating precession of the macroscopic magnetization by the application of a direct current \cite{Demidov2012,Duan2014,Demidov2014,Zholud2014,Collet2015}. Such a device, a spin-Hall oscillator, is patterned in a simple current in-plane geometry and contains no tunnel barriers. The ability of the spin-Hall effect to apply a spin-current transverse to the charge-current enables the use of conducting magnetic materials as well as low-damping, insulating materials like yttrium-iron-garnet (YIG) \cite{Collet2015}. This is in contrast to conventional spin-torque nano-oscillators where a spin-polarized current is passed into a conducting ferromagnetic layer \cite{Kiselev2003}. Despite being attractive due to their geometric simplicity and flexibility in choice of magnetic materials, for spin-Hall oscillators to become competitive with conventional spin-torque nano-oscillators, sources of damping such as spin-pumping \cite{Tserkovnyak2002} and multi-magnon scattering via exchange magnons \cite{Suhl1957,Demidov2011} should be addressed in order to reduce the linewidth and increase the power generation efficiency. In this Letter we show that the current induced excitation of incoherent exchange magnons can be observed using a direct current measurement. We find a change in the sample resistance which correlates with the reduction of saturation magnetization taken from our spectroscopy measurements, similar to previously reported results \cite{Demidov2011}. Both phenomena are consistent within a simple model only utilizing the precession cone-angle of the auto-oscillatory exchange magnons.

The sample is a Pt(3)/Py(4)/AlO$_\text{x}$(1.6) trilayer deposited by d.c.\ magnetron sputtering onto a MgO substrate. Numbers in parentheses give thicknesses in nanometers. 
The multilayer is patterned into a $\unit[500]{nm}\times \unit[6]{\mu m}$ bar via electron-beam lithography and subsequent Ar ion milling. Adjacent evaporated Cr(5)/Au(50) pads serve as contact pads. 
When current is applied to this structure, the portion of the current flowing inside the heavy metal (platinum) generates an out-of-plane spin-current via the spin-Hall effect (SHE). This spin-current exerts a spin-transfer torque (STT) on the magnetization of the ferromagnetic layer (permalloy) \cite{Berger1996,Slonczewski1996}, which can oppose the magnetic damping, leading to a change in the Gilbert damping parameter $\alpha$ \cite{Ando2008,Liu2011}:
\begin{equation}
\label{equ_stt}
\alpha = \alpha_0+\frac{\hbar}{2e} \frac{j_\text{c,hm} \theta_\text{SH}\sin \phi }{\left(H_0+\frac{M_\text{eff}}{2} \right) \mu_0 M_s t_\text{fm}}.
\end{equation}
Here, $\alpha_0$ is the pristine damping parameter, $j_\text{c,hm}$ is the charge-current density inside the heavy metal layer and $\theta_\text{SH}$ is the spin-Hall angle describing the conversion of charge- into spin-current. $\phi$ is the in-plane angle between the charge-current and the saturated magnetization of the ferromagnet, $H_0$ is the externally applied direct magnetic field amplitude, $M_\text{eff}$ is the effective magnetization taking into account a reduction of the saturation magnetization $M_s$ by the out-of-plane surface anisotropy and $t_\text{fm}$ is the ferromagnetic layer thickness.
By choosing the right polarity of $j_c$ and $\phi$, the STT becomes anti-damping. This can lead to auto-oscillations of the magnetization precession if the anti-damping STT exceeds the intrinsic Gilbert damping $\alpha_0$ \cite{Berger1996,Slonczewski1996}, thus implying a negative damping. 

We divide the auto-oscillatory spin-wave spectrum into two categories, namely dipolar ($\lambda\approx l \gg l_\text{ex}$) and exchange modes ($\lambda \approx l_\text{ex}$) \cite{Bayer2006}. Here, $\lambda$ denotes the wavelength of the spin-wave mode, $l=\unit[500]{nm}$ the length scale of the macroscopic sample and $l_\text{ex}=\unit[5]{nm}$ the exchange length of permalloy.
Dipolar modes lead to a coherent macroscopic magnetization oscillation and therefore to a macroscopic microwave resistance oscillation via the anisotropic magnetoresistance (AMR) of permallloy. This mixes with the applied direct current to a microwave voltage which can be detected as microwave power \cite{Kiselev2003}.
In contrast, the spatially varying resistance oscillation due to  exchange modes average over the whole sample size resulting in no macroscopic microwave power generation. 
However, because of their large phase volume, exchange modes dominate the localized oscillations of magnetic moments leading to a reduction of the effective saturation magnetization \cite{Demidov2011}.

\begin{figure}%
\includegraphics[width=8.5cm]{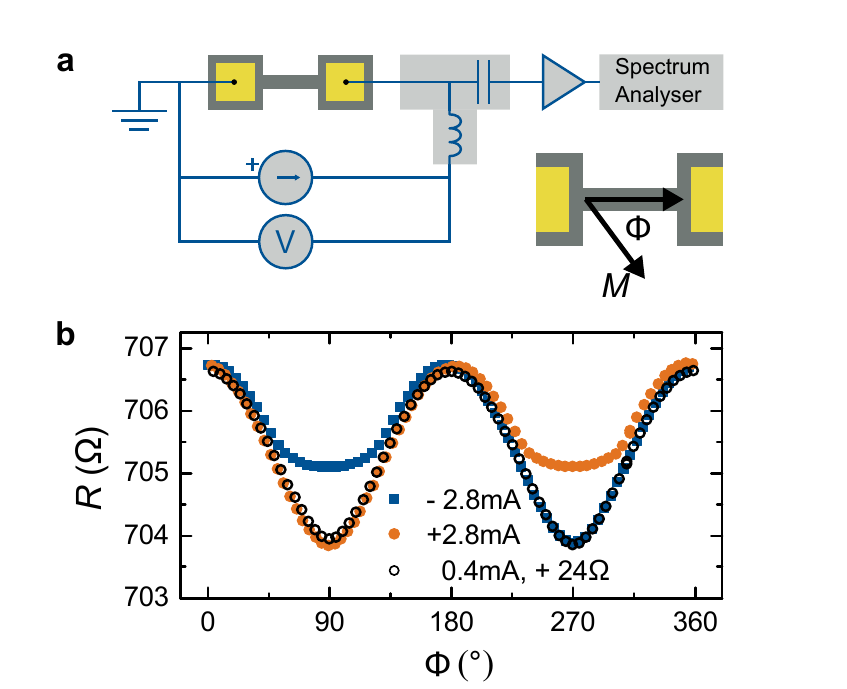}%
\caption{(a) Schematic of the experimental setup. A direct current source and voltmeter are connected to the nanowire via the low-frequency port of a bias-tee. The high-frequency port is connected to a microwave amplifier and commercial spectrum analyzer. The detail defines the in-plane angle $\phi$. Measurements are either performed at room-temperature (RT) or in a liquid helium bath at $\unit[2]{K}$ (LT). (b) RT-resistance versus in-plane angle $\phi$ at $\mu_0 H = \unit[55]{mT}$. 
A deviation from the expected $\cos^2\phi$ dependence (open circles) of anisotropic magnetoresistance is observed.}%
\label{fig_1}%
\end{figure}

We investigate in-plane saturated thin-films in which AMR only depends on the in-plane angle $\phi$ between the magnetization and the positive current direction:
\begin{equation}
R=R_\perp+(R_\parallel-R_\perp)\cos^2\phi=R_\perp+\Delta R \cos^2 \phi,
\end{equation} 
where $R_\perp$ and $R_\parallel$ are the resistances for perpendicular and parallel alignment respectively and $\Delta R$ is the magnetoresistance coefficient.
In order to simultaneously detect the generated microwave power as well as a change in resistance we use the circuitry depicted in Fig.~\ref{fig_1}a.

In the experiment, we rotate the in-plane magnetic field of fixed amplitude ($\unit[55]{mT}$) and record the room-temperature d.c.\ resistance, depicted in Fig.~\ref{fig_1}b. The resistance curves obtained at opposite current directions ($\pm\unit[2.8]{mA}$) exhibit an asymmetry, in contrast to the expected $\cos^2\phi$ dependence \cite{Thomson1857} observed for a smaller current amplitude ($\unit[0.4]{mA}$). The latter plot is off-set by $\unit[24]{\Omega}$ to compensate for Joule heating. As a constant current is applied, no temperature-effects beyond the offset are expected.
This asymmetry has been observed for more than five samples of different nanowire dimensions and for room-temperature as well as low-temperature ($T=\unit{2}{K}$) experiments. In the following, low-temperature data will be discussed as this allows the observation of coherent auto-oscillations \cite{Duan2014}.

\begin{figure}%
\includegraphics[width=8.5cm]{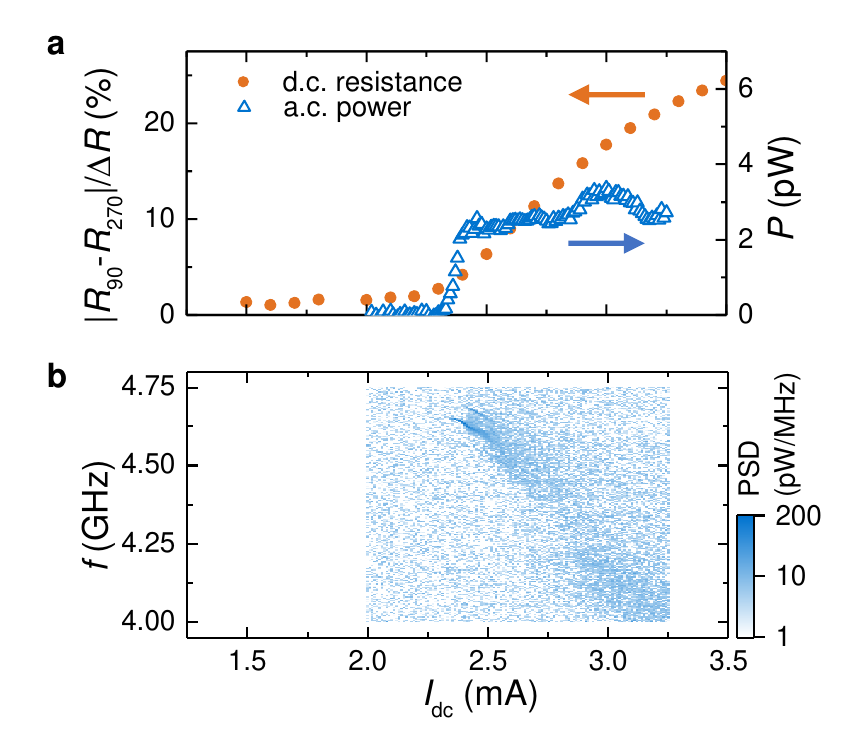}%
\caption{(a) Asymmetry of AMR normalized to anisotropic magnetoresistance as a function of applied direct current. In the same plot we show the integrated power taken from power spectral density measurements of coherent microwave auto-oscillations depicted in (b). Both measurements are performed at the same external magnetic field and at low temperature.
}
\label{fig_2}%
\end{figure}

We now investigate the current amplitude dependence of the observed asymmetry. 
In Fig.~\ref{fig_2}a, we plot the difference in resistance between $\phi=\unit[90]{^\circ}$ and $\phi=\unit[270]{^\circ}$ normalized to the magnetoresistive ratio $\Delta R$ of the sample at $\mu_0 H = \unit[55]{mT}$.
For small currents, the AMR is symmetric. Starting at $I_\text{dc}=\unit[2.3]{mA}$, the asymmetry evolves up to $\unit[26]{\%}$ at $I_\text{dc}=\unit[3.5]{mA}$, where it is still growing.
In the same plot we show the integrated microwave power emitted by the dipolar auto-oscillations measured via the microwave spectrum analyzer. 
The data is obtained at the same magnetic field at $\phi=\unit[260]{^\circ}$. 
 The integrated power clearly shows a threshold dependence with a sudden increase at $I_\text{dc}=\unit[2.35]{mA}$ after which it stays about constant at $P=\unit[2.5]{pW}$. These results show a correlation between the asymmetry in AMR and the microwave power generation via the spin-transfer torque induced auto-oscillation of magnetization.

In addition to the integrated power, we show its power spectral density in Fig.~\ref{fig_2}b, where the frequency resolution shows a red-shift and enhancement in linewidth with increasing direct current, all in accordance to previous results on microwave power generation by Duan et al.\cite{Duan2014}.
\begin{figure}%
\includegraphics[width=8.5cm]{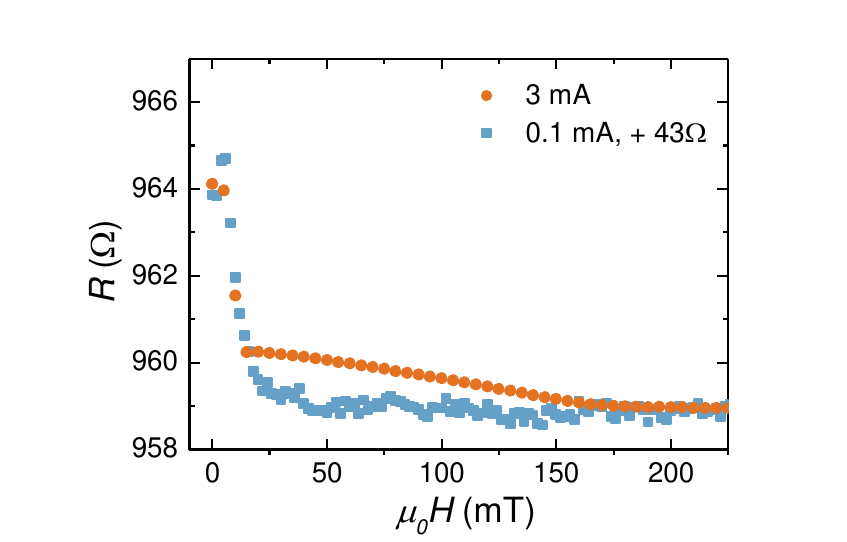}%
\caption{Field sweep of resistance along $\phi=\unit[270]{^\circ}$ at low temperature. For currents smaller than the introduced threshold current, the curve matches the expectation of AMR. For higher currents, an intermediate region is observed which qualitatively matches the asymmetry depicted in Fig.~\ref{fig_1}b.}%
\label{fig_3}%
\end{figure}

Another indication for the dynamic magnetization origin of the observed asymmetry is given in Fig.~\ref{fig_3}.
We perform field sweeps in-plane perpendicular to the bar with two different currents, one below the threshold current ($\unit[0.1]{mA}$) and one above ($\unit[3]{mA}$).
In the low-current graph, the magnetization saturates for external fields larger than $\unit[30]{mT}$, leading to a minimum in resistance due to AMR.
The high-current sweep, however, shows an intermediate regime where the resistance is still slightly larger. With increasing magnetic field the resistance declines towards its low-current value. No shift in the saturation field of permalloy of about $\unit[30]{mT}$ is observed and the asymmetry extends to a noticeably big magnetic field range. 
An additional Oersted field due to the increased current in the platinum layer would shift the field necessary for saturation, which can lead to an asymmetry in AMR if the sample reaches the unsaturated regime. Fig.~\ref{fig_3} shows that this is not the case for the observed effect.

We attribute the observed asymmetry in resistance to photoresistance of the spin-transfer torque induced auto-oscillatory exchange spin-wave modes.
Depending on the current direction, spin-transfer torque is anti-damping either at $\phi=\unit[90]{^\circ}$ or at $\phi=\unit[270]{^\circ}$ (Eq.~\eqref{equ_stt}). 
Thus, coherent auto-oscillations and exchange magnons are excited for only one of the directions, which qualitatively matches the asymmetry in d.c.\ resistance. Quantitatively, the threshold current for coherent dipolar auto-oscillations equals the threshold current for the discussed asymmetry. 
Not only in the current-domain, but also in the field-domain the theory of auto-oscillatory modes matches our observation. As $\alpha_0$ increases and $\Delta \alpha$ decreases with external field $H$, auto-oscillations remain only up to a certain field, in agreement with the field-dependence of asymmetry indicated in Fig.~\ref{fig_3}.
It is well known that coherent dipolar auto-oscillations can only sustain at low temperature \cite{Duan2014} due to the scattering with auto-oscillatory, thermally excited exchange magnons\cite{Demidov2011}.
Hence our observation of asymmetry at room-temperature further supports our explanation.

We model the excitation of exchange magnons as the oscillation of the magnetization of regions of size on the order of the exchange length of permalloy. These oscillations have spatially varying a.c.\ phases. Due to photoresistance \cite{Mecking2007}, the auto-oscillation leads to a change in d.c.\ resistance of each region. Summed over the whole sample, this change is given by 
\begin{equation}
\delta R_\text{dc} = -\frac{1}{2}\Delta R\phi^2_c\cos 2\phi,
\label{equ_photoresistance}
\end{equation}
where $\phi_c$ is the precessional cone-angle of exchange magnons, which we assume constant over the sample. 
Thus the exchange modes do not contribute to the microwave power generation due to the averaging of the a.c.\ phases, but their impact on the d.c.\ resistance can still be detected.

\begin{figure}%
\includegraphics[width=8.5cm]{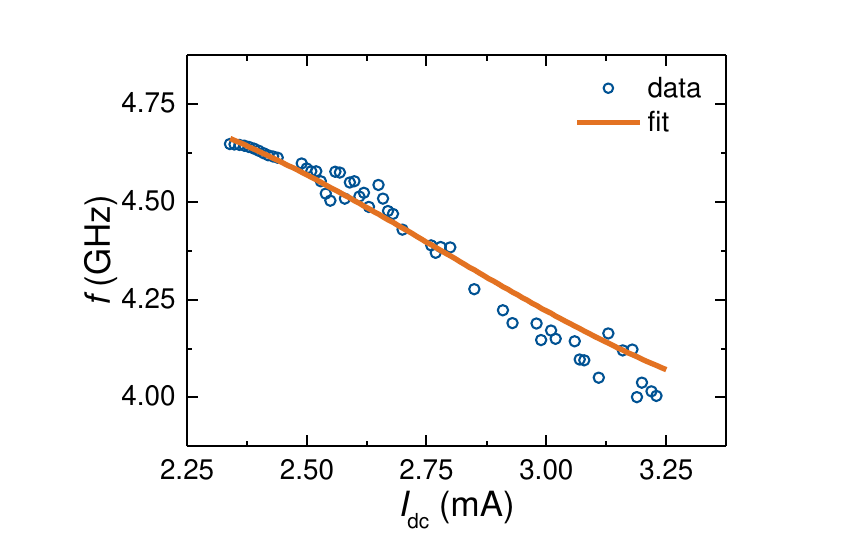}%
\caption{Resonance frequency versus applied bias current extracted from Fig.~\ref{fig_2}b. The fit according to the model of exchange magnons given in the main text matches the observed data.}%
\label{fig_4}%
\end{figure}

In addition to photoresistance, exchange magnons manifest themselves in the reduction of saturation magnetization as has been reported before in Brillouin light scattering experiments \cite{Demidov2011}.
We combine the reduction in saturation magnetization and asymmetry in d.c.\ resistance within the introduced model. Using the asymmetry in AMR taken from Fig.~\ref{fig_2}a, we calculate the exchange mode precession angle $\phi_c$ via Eq.~\eqref{equ_photoresistance}. We derive a reduced effective saturation magnetization by taking the projection of the magnetization precession onto its equilibrium direction, giving $M_\text{eff,red}=M_\text{eff} \cdot \cos{\phi_c}$.
This allows us to evaluate the resonance frequency of the emitted microwave power given by Kittel's equation as a function of applied direct current. By compensating for the additional current-induced Oersted field, we obtain Fig.~\ref{fig_4}, in which the resonance frequency extracted from Fig.~\ref{fig_2}b is plotted as a function of applied direct current. 
The line is a fit according to the simple model explained above. We use an anisotropy field of $H_a=\unit[10]{mT}$ estimated as the demagnetization field in the bar with the given dimensions, and zero current effective saturation magnetization of $M_\text{eff}=\unit[0.71]{T}$, extracted from ferromagnetic resonance studies. Additional contributions to the change of saturation magnetization due to Joule heating have been found to be negligible compared to the impact of STT even at room-temperature \cite{Demidov2011}, where the change in saturation magnetization due to temperature fluctuations is more pronounced compared to the presented low temperature study following Bloch's $T^{3/2}$-law.
Figure~\ref{fig_4} thus nicely shows how the excitation of exchange magnons plays a crucial role for a.c.\ and d.c.\ properties of spin-Hall oscillators.

In conclusion, we have shown the importance of auto-oscillatory exchange spin-wave modes in spin-Hall oscillators, which manifest themselves in the resonance frequency of microwave power generation as well as the d.c.\ resistance. 
Our findings allow the study of exchange spin-wave modes in spin-Hall oscillators without the need for either low-temperature or high-frequency setups as they conveniently appear in the most basic property of electronics, the resistance. This should pave the way for further understanding of the excitation and necessary suppression of exchange spin-waves to increase the efficiency of spin-Hall oscillators.

We are grateful for useful discussions with Chiara Ciccarelli and James Haigh. This project was partly funded by EPSRC under EP/M50693X/1 and EP/L027151/1. A.J.F. is supported by ERC Grant No. 648613 and a Hitachi research fellowship.
\bibliography{DCbib}

\begin{thebibliography}{16}%
\makeatletter
\providecommand \@ifxundefined [1]{%
 \@ifx{#1\undefined}
}%
\providecommand \@ifnum [1]{%
 \ifnum #1\expandafter \@firstoftwo
 \else \expandafter \@secondoftwo
 \fi
}%
\providecommand \@ifx [1]{%
 \ifx #1\expandafter \@firstoftwo
 \else \expandafter \@secondoftwo
 \fi
}%
\providecommand \natexlab [1]{#1}%
\providecommand \enquote  [1]{``#1''}%
\providecommand \bibnamefont  [1]{#1}%
\providecommand \bibfnamefont [1]{#1}%
\providecommand \citenamefont [1]{#1}%
\providecommand \href@noop [0]{\@secondoftwo}%
\providecommand \href [0]{\begingroup \@sanitize@url \@href}%
\providecommand \@href[1]{\@@startlink{#1}\@@href}%
\providecommand \@@href[1]{\endgroup#1\@@endlink}%
\providecommand \@sanitize@url [0]{\catcode `\\12\catcode `\$12\catcode
  `\&12\catcode `\#12\catcode `\^12\catcode `\_12\catcode `\%12\relax}%
\providecommand \@@startlink[1]{}%
\providecommand \@@endlink[0]{}%
\providecommand \url  [0]{\begingroup\@sanitize@url \@url }%
\providecommand \@url [1]{\endgroup\@href {#1}{\urlprefix }}%
\providecommand \urlprefix  [0]{URL }%
\providecommand \Eprint [0]{\href }%
\providecommand \doibase [0]{http://dx.doi.org/}%
\providecommand \selectlanguage [0]{\@gobble}%
\providecommand \bibinfo  [0]{\@secondoftwo}%
\providecommand \bibfield  [0]{\@secondoftwo}%
\providecommand \translation [1]{[#1]}%
\providecommand \BibitemOpen [0]{}%
\providecommand \bibitemStop [0]{}%
\providecommand \bibitemNoStop [0]{.\EOS\space}%
\providecommand \EOS [0]{\spacefactor3000\relax}%
\providecommand \BibitemShut  [1]{\csname bibitem#1\endcsname}%
\let\auto@bib@innerbib\@empty
\bibitem [{\citenamefont {Demidov}\ \emph {et~al.}(2012)\citenamefont
  {Demidov}, \citenamefont {Urazhdin}, \citenamefont {Ulrichs}, \citenamefont
  {Tiberkevich}, \citenamefont {Slavin}, \citenamefont {Baither}, \citenamefont
  {Schmitz},\ and\ \citenamefont {Demokritov}}]{Demidov2012}%
  \BibitemOpen
  \bibfield  {author} {\bibinfo {author} {\bibfnamefont {V.~E.}\ \bibnamefont
  {Demidov}}, \bibinfo {author} {\bibfnamefont {S.}~\bibnamefont {Urazhdin}},
  \bibinfo {author} {\bibfnamefont {H.}~\bibnamefont {Ulrichs}}, \bibinfo
  {author} {\bibfnamefont {V.}~\bibnamefont {Tiberkevich}}, \bibinfo {author}
  {\bibfnamefont {A.}~\bibnamefont {Slavin}}, \bibinfo {author} {\bibfnamefont
  {D.}~\bibnamefont {Baither}}, \bibinfo {author} {\bibfnamefont
  {G.}~\bibnamefont {Schmitz}}, \ and\ \bibinfo {author} {\bibfnamefont
  {S.~O.}\ \bibnamefont {Demokritov}},\ }\href {\doibase 10.1038/nmat3459}
  {\bibfield  {journal} {\bibinfo  {journal} {Nature materials}\ }\textbf
  {\bibinfo {volume} {11}},\ \bibinfo {pages} {1028} (\bibinfo {year}
  {2012})}\BibitemShut {NoStop}%
\bibitem [{\citenamefont {Duan}\ \emph {et~al.}(2014)\citenamefont {Duan},
  \citenamefont {Smith}, \citenamefont {Yang}, \citenamefont {Youngblood},
  \citenamefont {Lindner}, \citenamefont {Demidov}, \citenamefont
  {Demokritov},\ and\ \citenamefont {Krivorotov}}]{Duan2014}%
  \BibitemOpen
  \bibfield  {author} {\bibinfo {author} {\bibfnamefont {Z.}~\bibnamefont
  {Duan}}, \bibinfo {author} {\bibfnamefont {A.}~\bibnamefont {Smith}},
  \bibinfo {author} {\bibfnamefont {L.}~\bibnamefont {Yang}}, \bibinfo {author}
  {\bibfnamefont {B.}~\bibnamefont {Youngblood}}, \bibinfo {author}
  {\bibfnamefont {J.}~\bibnamefont {Lindner}}, \bibinfo {author} {\bibfnamefont
  {V.~E.}\ \bibnamefont {Demidov}}, \bibinfo {author} {\bibfnamefont {S.~O.}\
  \bibnamefont {Demokritov}}, \ and\ \bibinfo {author} {\bibfnamefont {I.~N.}\
  \bibnamefont {Krivorotov}},\ }\href {\doibase 10.1038/ncomms6616} {\bibfield
  {journal} {\bibinfo  {journal} {Nature Communications}\ }\textbf {\bibinfo
  {volume} {5}},\ \bibinfo {pages} {5616} (\bibinfo {year} {2014})}\BibitemShut
  {NoStop}%
\bibitem [{\citenamefont {Demidov}\ \emph {et~al.}(2014)\citenamefont
  {Demidov}, \citenamefont {Urazhdin}, \citenamefont {Zholud}, \citenamefont
  {Sadovnikov},\ and\ \citenamefont {Demokritov}}]{Demidov2014}%
  \BibitemOpen
  \bibfield  {author} {\bibinfo {author} {\bibfnamefont {V.~E.}\ \bibnamefont
  {Demidov}}, \bibinfo {author} {\bibfnamefont {S.}~\bibnamefont {Urazhdin}},
  \bibinfo {author} {\bibfnamefont {A.}~\bibnamefont {Zholud}}, \bibinfo
  {author} {\bibfnamefont {a.~V.}\ \bibnamefont {Sadovnikov}}, \ and\ \bibinfo
  {author} {\bibfnamefont {S.~O.}\ \bibnamefont {Demokritov}},\ }\href
  {\doibase 10.1063/1.4901027} {\bibfield  {journal} {\bibinfo  {journal}
  {Applied Physics Letters}\ }\textbf {\bibinfo {volume} {105}},\ \bibinfo
  {pages} {172410} (\bibinfo {year} {2014})}\BibitemShut {NoStop}%
\bibitem [{\citenamefont {Zholud}\ and\ \citenamefont
  {Urazhdin}(2014)}]{Zholud2014}%
  \BibitemOpen
  \bibfield  {author} {\bibinfo {author} {\bibfnamefont {A.}~\bibnamefont
  {Zholud}}\ and\ \bibinfo {author} {\bibfnamefont {S.}~\bibnamefont
  {Urazhdin}},\ }\href {\doibase 10.1063/1.4896023} {\bibfield  {journal}
  {\bibinfo  {journal} {Applied Physics Letters}\ }\textbf {\bibinfo {volume}
  {105}},\ \bibinfo {pages} {112404} (\bibinfo {year} {2014})}\BibitemShut
  {NoStop}%
\bibitem [{\citenamefont {Collet}\ \emph {et~al.}(2015)\citenamefont {Collet},
  \citenamefont {Milly}, \citenamefont {Kelly}, \citenamefont {Naletov},
  \citenamefont {Bernard}, \citenamefont {Bortolotti}, \citenamefont {Demidov},
  \citenamefont {Demokritov}, \citenamefont {Prieto}, \citenamefont {Mu},
  \citenamefont {Cros}, \citenamefont {Anane}, \citenamefont {Loubens},\ and\
  \citenamefont {Klein}}]{Collet2015}%
  \BibitemOpen
  \bibfield  {author} {\bibinfo {author} {\bibfnamefont {M.}~\bibnamefont
  {Collet}}, \bibinfo {author} {\bibfnamefont {X.~D.}\ \bibnamefont {Milly}},
  \bibinfo {author} {\bibfnamefont {O.~A.}\ \bibnamefont {Kelly}}, \bibinfo
  {author} {\bibfnamefont {V.~V.}\ \bibnamefont {Naletov}}, \bibinfo {author}
  {\bibfnamefont {R.}~\bibnamefont {Bernard}}, \bibinfo {author} {\bibfnamefont
  {P.}~\bibnamefont {Bortolotti}}, \bibinfo {author} {\bibfnamefont {V.~E.}\
  \bibnamefont {Demidov}}, \bibinfo {author} {\bibfnamefont {O.}~\bibnamefont
  {Demokritov}}, \bibinfo {author} {\bibfnamefont {J.~L.}\ \bibnamefont
  {Prieto}}, \bibinfo {author} {\bibfnamefont {M.}~\bibnamefont {Mu}}, \bibinfo
  {author} {\bibfnamefont {V.}~\bibnamefont {Cros}}, \bibinfo {author}
  {\bibfnamefont {A.}~\bibnamefont {Anane}}, \bibinfo {author} {\bibfnamefont
  {G.~D.}\ \bibnamefont {Loubens}}, \ and\ \bibinfo {author} {\bibfnamefont
  {O.}~\bibnamefont {Klein}},\ }\href {\doibase arXiv:1504:01512} {\bibfield
  {journal} {\bibinfo  {journal} {arXiv:1504.01512}\ ,\ \bibinfo {pages} {1}}
  (\bibinfo {year} {2015})}\BibitemShut {NoStop}%
\bibitem [{\citenamefont {Kiselev}, \citenamefont {Sankey},\ and\ \citenamefont
  {Krivorotov}(2003)}]{Kiselev2003}%
  \BibitemOpen
  \bibfield  {author} {\bibinfo {author} {\bibfnamefont {S.~I.}\ \bibnamefont
  {Kiselev}}, \bibinfo {author} {\bibfnamefont {J.~C.}\ \bibnamefont {Sankey}},
  \ and\ \bibinfo {author} {\bibfnamefont {I.~N.}\ \bibnamefont {Krivorotov}},\
  }\href {\doibase 10.1038/nature01967} {\bibfield  {journal} {\bibinfo
  {journal} {Nature}\ }\textbf {\bibinfo {volume} {425}},\ \bibinfo {pages}
  {1207} (\bibinfo {year} {2003})}\BibitemShut {NoStop}%
\bibitem [{\citenamefont {Tserkovnyak}, \citenamefont {Brataas},\ and\
  \citenamefont {Bauer}(2002)}]{Tserkovnyak2002}%
  \BibitemOpen
  \bibfield  {author} {\bibinfo {author} {\bibfnamefont {Y.}~\bibnamefont
  {Tserkovnyak}}, \bibinfo {author} {\bibfnamefont {A.}~\bibnamefont
  {Brataas}}, \ and\ \bibinfo {author} {\bibfnamefont {G.~E.~W.}\ \bibnamefont
  {Bauer}},\ }\href {\doibase 10.1103/PhysRevLett.88.117601} {\bibfield
  {journal} {\bibinfo  {journal} {Physical Review Letters}\ }\textbf {\bibinfo
  {volume} {88}},\ \bibinfo {pages} {117601} (\bibinfo {year}
  {2002})}\BibitemShut {NoStop}%
\bibitem [{\citenamefont {Suhl}(1957)}]{Suhl1957}%
  \BibitemOpen
  \bibfield  {author} {\bibinfo {author} {\bibfnamefont {H.}~\bibnamefont
  {Suhl}},\ }\href {\doibase 10.1016/0022-3697(57)90010-0} {\bibfield
  {journal} {\bibinfo  {journal} {Journal of Physics and Chemistry of Solids}\
  }\textbf {\bibinfo {volume} {1}},\ \bibinfo {pages} {209} (\bibinfo {year}
  {1957})}\BibitemShut {NoStop}%
\bibitem [{\citenamefont {Demidov}\ \emph {et~al.}(2011)\citenamefont
  {Demidov}, \citenamefont {Urazhdin}, \citenamefont {Edwards}, \citenamefont
  {Stiles}, \citenamefont {McMichael},\ and\ \citenamefont
  {Demokritov}}]{Demidov2011}%
  \BibitemOpen
  \bibfield  {author} {\bibinfo {author} {\bibfnamefont {V.~E.}\ \bibnamefont
  {Demidov}}, \bibinfo {author} {\bibfnamefont {S.}~\bibnamefont {Urazhdin}},
  \bibinfo {author} {\bibfnamefont {E.~R.~J.}\ \bibnamefont {Edwards}},
  \bibinfo {author} {\bibfnamefont {M.~D.}\ \bibnamefont {Stiles}}, \bibinfo
  {author} {\bibfnamefont {R.~D.}\ \bibnamefont {McMichael}}, \ and\ \bibinfo
  {author} {\bibfnamefont {S.~O.}\ \bibnamefont {Demokritov}},\ }\href
  {\doibase 10.1103/PhysRevLett.107.107204} {\bibfield  {journal} {\bibinfo
  {journal} {Physical Review Letters}\ }\textbf {\bibinfo {volume} {107}},\
  \bibinfo {pages} {107204} (\bibinfo {year} {2011})}\BibitemShut {NoStop}%
\bibitem [{\citenamefont {Berger}(1996)}]{Berger1996}%
  \BibitemOpen
  \bibfield  {author} {\bibinfo {author} {\bibfnamefont {L.}~\bibnamefont
  {Berger}},\ }\href {\doibase 10.1103/PhysRevB.54.9353} {\bibfield  {journal}
  {\bibinfo  {journal} {Physical Review B}\ }\textbf {\bibinfo {volume} {54}},\
  \bibinfo {pages} {9353} (\bibinfo {year} {1996})}\BibitemShut {NoStop}%
\bibitem [{\citenamefont {Slonczewski}(1996)}]{Slonczewski1996}%
  \BibitemOpen
  \bibfield  {author} {\bibinfo {author} {\bibfnamefont {J.~C.}\ \bibnamefont
  {Slonczewski}},\ }\href {\doibase 10.1016/0304-8853(96)00062-5} {\bibfield
  {journal} {\bibinfo  {journal} {Journal of Magnetism and Magnetic Materials}\
  }\textbf {\bibinfo {volume} {159}},\ \bibinfo {pages} {1} (\bibinfo {year}
  {1996})}\BibitemShut {NoStop}%
\bibitem [{\citenamefont {Ando}\ \emph {et~al.}(2008)\citenamefont {Ando},
  \citenamefont {Takahashi}, \citenamefont {Harii}, \citenamefont {Sasage},
  \citenamefont {Ieda}, \citenamefont {Maekawa},\ and\ \citenamefont
  {Saitoh}}]{Ando2008}%
  \BibitemOpen
  \bibfield  {author} {\bibinfo {author} {\bibfnamefont {K.}~\bibnamefont
  {Ando}}, \bibinfo {author} {\bibfnamefont {S.}~\bibnamefont {Takahashi}},
  \bibinfo {author} {\bibfnamefont {K.}~\bibnamefont {Harii}}, \bibinfo
  {author} {\bibfnamefont {K.}~\bibnamefont {Sasage}}, \bibinfo {author}
  {\bibfnamefont {J.}~\bibnamefont {Ieda}}, \bibinfo {author} {\bibfnamefont
  {S.}~\bibnamefont {Maekawa}}, \ and\ \bibinfo {author} {\bibfnamefont
  {E.}~\bibnamefont {Saitoh}},\ }\href {\doibase
  10.1103/PhysRevLett.101.036601} {\bibfield  {journal} {\bibinfo  {journal}
  {Physical Review Letters}\ }\textbf {\bibinfo {volume} {101}},\ \bibinfo
  {pages} {036601} (\bibinfo {year} {2008})}\BibitemShut {NoStop}%
\bibitem [{\citenamefont {Liu}\ \emph {et~al.}(2011)\citenamefont {Liu},
  \citenamefont {Moriyama}, \citenamefont {Ralph},\ and\ \citenamefont
  {Buhrman}}]{Liu2011}%
  \BibitemOpen
  \bibfield  {author} {\bibinfo {author} {\bibfnamefont {L.}~\bibnamefont
  {Liu}}, \bibinfo {author} {\bibfnamefont {T.}~\bibnamefont {Moriyama}},
  \bibinfo {author} {\bibfnamefont {D.~C.}\ \bibnamefont {Ralph}}, \ and\
  \bibinfo {author} {\bibfnamefont {R.~A.}\ \bibnamefont {Buhrman}},\ }\href
  {\doibase 10.1103/PhysRevLett.106.036601} {\bibfield  {journal} {\bibinfo
  {journal} {Physical Review Letters}\ }\textbf {\bibinfo {volume} {106}},\
  \bibinfo {pages} {036601} (\bibinfo {year} {2011})}\BibitemShut {NoStop}%
\bibitem [{\citenamefont {Bayer}\ \emph {et~al.}(2006)\citenamefont {Bayer},
  \citenamefont {Jorzick}, \citenamefont {Demokritov}, \citenamefont {Slavin},
  \citenamefont {Guslienko}, \citenamefont {Berkov}, \citenamefont {Gorn},
  \citenamefont {Kostylev},\ and\ \citenamefont {Hillebrands}}]{Bayer2006}%
  \BibitemOpen
  \bibfield  {author} {\bibinfo {author} {\bibfnamefont {C.}~\bibnamefont
  {Bayer}}, \bibinfo {author} {\bibfnamefont {J.}~\bibnamefont {Jorzick}},
  \bibinfo {author} {\bibfnamefont {S.~O.}\ \bibnamefont {Demokritov}},
  \bibinfo {author} {\bibfnamefont {A.~N.}\ \bibnamefont {Slavin}}, \bibinfo
  {author} {\bibfnamefont {K.~Y.}\ \bibnamefont {Guslienko}}, \bibinfo {author}
  {\bibfnamefont {D.~V.}\ \bibnamefont {Berkov}}, \bibinfo {author}
  {\bibfnamefont {N.~L.}\ \bibnamefont {Gorn}}, \bibinfo {author}
  {\bibfnamefont {M.~P.}\ \bibnamefont {Kostylev}}, \ and\ \bibinfo {author}
  {\bibfnamefont {B.}~\bibnamefont {Hillebrands}},\ }\href {\doibase
  10.1007/10938171\_2} {\bibfield  {journal} {\bibinfo  {journal} {Topics in
  Applied Physics}\ }\textbf {\bibinfo {volume} {101}},\ \bibinfo {pages} {57}
  (\bibinfo {year} {2006})}\BibitemShut {NoStop}%
\bibitem [{\citenamefont {Thomson}(1857)}]{Thomson1857}%
  \BibitemOpen
  \bibfield  {author} {\bibinfo {author} {\bibfnamefont {W.}~\bibnamefont
  {Thomson}},\ }\href {\doibase 10.1098/rspl.1902.0058} {\bibfield  {journal}
  {\bibinfo  {journal} {Proceedings of the Royal Society of London}\ }\textbf
  {\bibinfo {volume} {8}},\ \bibinfo {pages} {546} (\bibinfo {year}
  {1857})}\BibitemShut {NoStop}%
\bibitem [{\citenamefont {Mecking}, \citenamefont {Gui},\ and\ \citenamefont
  {Hu}(2007)}]{Mecking2007}%
  \BibitemOpen
  \bibfield  {author} {\bibinfo {author} {\bibfnamefont {N.}~\bibnamefont
  {Mecking}}, \bibinfo {author} {\bibfnamefont {Y.~S.}\ \bibnamefont {Gui}}, \
  and\ \bibinfo {author} {\bibfnamefont {C.~M.}\ \bibnamefont {Hu}},\ }\href
  {\doibase 10.1103/PhysRevB.76.224430} {\bibfield  {journal} {\bibinfo
  {journal} {Physical Review B - Condensed Matter and Materials Physics}\
  }\textbf {\bibinfo {volume} {76}},\ \bibinfo {pages} {1} (\bibinfo {year}
  {2007})}\BibitemShut {NoStop}%
\end{thebibliography}%

\end{document}